\documentclass{article}

\usepackage{arxiv}

\usepackage[utf8]{inputenc} 
\usepackage[T1]{fontenc}    
\usepackage{hyperref}       
\usepackage{url}            
\usepackage{booktabs}       
\usepackage{amsfonts}       
\usepackage{nicefrac}       
\usepackage{lipsum}		
\usepackage{natbib}
\usepackage{doi}

\usepackage[pdftex]{graphicx} 
\usepackage[ruled,vlined]{algorithm2e}
\usepackage{setspace}
\usepackage{pdfpages}

\usepackage[all]{nowidow} 
\usepackage[protrusion=true,expansion=true]{microtype} 
\usepackage{amsmath}
\usepackage{amssymb}

\DeclareMathOperator*{\argmax}{arg\,max}
\usepackage{amsthm}
\newtheorem{theorem}{Theorem}
\theoremstyle{definition}

\allowdisplaybreaks[1]
\newcommand\numberthis{\addtocounter{equation}{1}\tag{\theequation}}
\usepackage[thinc]{esdiff}
\usepackage{bbm}

\usepackage{caption}
\usepackage{float}
\usepackage[caption = false]{subfig}
\usepackage{listings}
\usepackage{xcolor}
\usepackage{listings}
\usepackage{etoolbox}

\renewenvironment{proof}{{\bfseries Proof}}{\qed}

\emergencystretch=\maxdimen
\title{Sequential Stochastic Network Structure Optimization With Applications To Addressing Canada’s Obesity Epidemic}

\date{} 					

\author{Nicholas A. G. Johnson \\
	Operations Research and Financial Engineering\\
	Princeton University\\
	Princeton, NJ 08544 \\
	\texttt{nagj@alumni.princeton.edu} \\
}

\renewcommand{\headeright}{}
\renewcommand{\undertitle}{}
\renewcommand{\shorttitle}{Sequential Stochastic Network Structure Optimization}

\hypersetup{
pdftitle={Sequential Stochastic Network Structure Optimization With Applications To Addressing Canada’s Obesity Epidemic},
pdfauthor={Nicholas A. G. Johnson},
}

\begin{document}
\maketitle

\begin{abstract}
	In this work, we introduce a novel mathematical network model for community level preventative health interventions. We develop algorithms to approximately solve this novel formulation at large scale and we rigorously explore their theoretical properties. We create a realistic simulation environment for interventions designed to curb the prevalence of obesity occurring in the region of Montreal, Canada, and use this environment to empirically evaluate the performance of the algorithms we develop. We find that our algorithms significantly outperform all baseline interventions. Moreover, for fixed computational resources, our algorithms address problems of significantly greater size than the best existing alternative algorithm.
\end{abstract}


\section{Introduction}    

Obesity is a chronic health condition defined as “abnormal or excessive fat accumulation that presents a risk to health” \cite{WHO}. Obesity must be taken very seriously; obese individuals experience greater risk of developing Type 2 diabetes and cardiovascular disease. According to the Obesity in Canada Report which was released on June 20, 2011, roughly 25\% of all Canadian adults are obese based on data taken from 2007 to 2009. From 1981 to 2009, obesity in the Canadian population has roughly doubled and continues to increase.

We focus on a community-based health intervention in which obese individuals are paired with healthy individuals in a mentor-mentee style relationship to encourage the adoption of healthy practices to curb obesity. Obesity is a condition that is heavily influenced by lifestyle choices (particularly physical activity and diet). A person’s lifestyle choices are often influenced relatively significantly by the choices of those they interact with most frequently \cite{soc_learn}. With this in mind, it is understandable that introducing additional healthy individuals into the social network of an obese person can have the effect of encouraging the obese individual to increase adoption of healthy practices to improve their health \cite{change_behavior}.

In this work, we focus on the mathematical formulation of the intervention described in the previous paragraph and on the algorithmic challenge of determining the ideal mentor-mentee pairings for large social networks efficiently. Concretely, we make 4 contributions:

\begin{enumerate}
    \item We propose a novel mathematical network model for preventative health interventions that allows for a dynamic intervention and that more appropriately models how humans change their behavior in response to individuals they interact with when compared to existing alternatives.
    \item We develop efficient algorithms to approximately solve the novel problem formulation at large scale.
    \item We rigorously explore the theoretical properties of the algorithms we develop.
    \item We create a realistic simulation environment of interventions occurring in the region of Montreal, Canada, informed by demographic and health data obtained from the 2015/2016 Canadian Community Health Survey \cite{stats_can_1}. We use this environment to empirically evaluate the performance of the algorithms we develop and find that they scale significantly better than an appropriately customised version of the best available alternative algorithm.
\end{enumerate}

\section{Problem Statement and Formulation}

\subsection{Modeling Framework}

In this section, the preventative health intervention described towards the end of the previous section is formalized mathematically. This formulation is inspired by the Preventative Health Network Intervention (PHNI) problem \cite{Wilder}. However, in this case, the preventative health intervention being considered is a sequential decision problem. Accordingly, the canonical sequential decision framework presented by Professor Warren Powell in \cite{RLSO} will be employed.

\subsection{State Variable}

The state variable $S_t$ contains all information needed to model the system from time $t$ onward. We are given a directed graph $G_0=(V,E_0)$ where vertices represent individuals in the community. A directed edge $(u,v)$ exists in $G$ if and only if agent $u$ exerts an influence on agent $v$. Moreover, each edge $(u,v) \in E_0$ has an associated weight $w_{(u,v)}(0) \in [0,1]$ which quantifies the influence agent $u$ exerts on agent $v$ relative to all agents that influence agent $v$. Let $\delta^{in}(v)$ denote all of agent $v$'s in neighbors. We have $\sum_{u \in \delta^{in}(v)} w_{(u,v)}(0) = 1$ $\forall v \in V$. Each node $v$ in the graph has a state $x_v(0) \in [0,1]$ which represents the health of agent $v$. Specifically, an individual who is obese would have value $0$ while an individual who is perfectly healthy would have value $1$. The set of edges in the graph, the edge weights and the health of each node change dynamically as the system evolves. At time $t$, these quantities are represented by $E_t$, $w_{(u,v)}(t)$ and $x_v(t)$. Thus, the state variable is given by \[S_t=(G_t=(V,E_t),\{w_{(u,v)}(t) : (u,v) \in E_t\},\{x_v(t) : v \in V\})\]

\subsection{Decision Variable}

The decision variable $X_t$ consists of the action that the agent takes in order to impact the system. There is some target set of nodes $S \subset V$. Think of this as the group of obese individuals who opted-in to the health program. In addition to the set of target nodes $S$, there is a set of healthy nodes $H \subset V$ which represents healthy individuals who have opted to participate in the intervention. Let \[\overline{E_t} = \{(u,v) : u \in H, v \in S, (u,v) \notin E_t\}\] $\overline{E_t}$ is the set of all edges from a healthy individual in $H$ to an individual in the target set $S$ that do not already exist in $G_t$. At time $t$, the agent selects $A_t \subseteq \overline{E_t}$. The set $A_t$ are edges that will be added to the graph at time $t+1$. We enforce the constraint that for any node, the number of edges added to that node at time $t$ can be at most equal to the in-degree of that node. Define $\chi_v^{in}(A_t) \, \forall \, v \in V$ as follows: \[\chi_v^{in}(A_t) = |\{(x,y) \in A_t : y = v\}|\] $\chi_v^{in}(A_t)$ represents the number of in-going edges added to node $v$ as a result of taking decision $A_t$. The decision variable $X_t$ is an element in the set \[D_t = \{a_t \, | \, a_t \subseteq \overline{E_t}, \, \chi_v^{in}(a_t) \leq |\delta^{in}(v)| \,\,\, \forall \, v \in V \} \]

\subsection{Exogenous Information}

Exogenous information $I_t$ encapsulates all non-deterministic aspects of the system. Whenever a new edge $(u,v)$ is introduced into the graph, an edge $(w,v)$ for some $w \in \delta^{in}(v)$ must be removed from the graph. This models the fact that an individual can only be influenced socially by a limited number of individuals. Define $E_t(A_t)$ as follows: \[E_t(A_t) = \{R \subseteq E_t : rem_v(R) = \chi_v^{in}(A_t) \, \forall \, v \in V\}\] where $rem_v(R) = |\{(x,y) \in R : y = v\}|$. Intuitively, $E_t(A_t)$ represents the collection of all sets of edges in $E_t$ satisfying the condition that if those edges are removed and the edges in $A_t$ are added, the in-degree of any node $v$ will be unchanged. Edge removal in response to added edges occurs randomly as follows: $\, \forall \, v \in V$ choose $\chi_v^{in}(A_t)$ edges randomly from the set $\{(u,v) : u \in \delta^{in}(v)\}$ and include them in $R_t$ according to the probability distribution given by $\mathbb{P}\big{[}(r, v) \in R_t\big{]} = \frac{1-w_{(r, v)}(t)}{|\delta^{in}(v)|-1}$. We refer to this process as weighted edge removals. This models the fact that people are less likely to lose contact with individuals who exert strong social influence on them than they are to lose contact with those who exert little social influence on them.

\subsection{System Dynamics}

The system dynamics specify how the system evolves from time $t$ to time $t+1$ after the agent takes action $A_t$ and observes exogenous information $I_t$. Specifically, the system dynamics can be thought of as a transition function $S^M$ that maps $S_t, A_t$ and $I_t$ to $S_{t+1}$ (in other  words, $S_{t+1} = S^M(S_t, A_t, I_t)$). Thus, the system dynamics must specify $E_{t+1}, \,\, w_{(u,v)}(t+1) \, \forall \, (u,v) \in E_{t+1}$ and $x_v(t+1) \, \forall \, v \in V$. $E_{t+1}$ is updated in a straightforward manner given $A_t$ and $R_t$: \[E_{t+1} = (E_t \setminus R_t) \cup A_t\] Note that due to the definition of $R_t$ and $A_t$, the in-degree of each node v is constant at all times $t$. Edge weights will evolve over time towards placing equal weight on all in-neighbors to a given node. The rate of this evolution will be governed by a parameter of the system $\mu \in (0,1)$. When a new edge is introduced into the graph, it will be initialized to some small value to model the fact that an individual initially will not significantly trust a new individual introduced into their immediate social network. Weight updates will occur according to Algorithm 1. Letting $g$ denote Algorithm 1, we have \[\{w_{(u,v)}(t+1) : (u,v) \in E_{t+1}\} = g(\{w_{(u,v)}(t) : (u,v) \in E_{t}\}, \, A_t)\]

\begin{algorithm}
\setstretch{1}
\SetAlgoLined
\KwResult{$w_{(u,v)}(t+1) \, \forall \, (u,v) \in E_{t+1}$}
 \For{$v \in V$}{
  \eIf{$\chi_v^{in}(A_t)=0$}{  
    \For{$u \in \delta^{in}(v)$}{
      $w_{(u,v)}(t+1) = (1-\mu)w_{(u,v)}(t)+ \frac{\mu}{|\delta^{in}(v)|}$
    }
  }{
    \For{$u \in \delta^{in}(v)$}{
      \eIf{$(u,v) \in A_t$}{
        $w_{(u,v)}(t+1) = \frac{1}{|\delta^{in}(v)|^2}$
      }{
        $w_{(u,v)}(t+1) = w_{(u,v)}(t)$
      }
    }
    \For{$u \in \delta^{in}(v)$}{
      $w_{(u,v)}(t+1) = \frac{w_{(u,v)}(t+1)}{\sum_{u \in \delta^{in}(v)} w_{(u,v)}(t+1)}$
    }
  }
}
 \caption{Edge Weight Updates}
\end{algorithm}
 
\noindent
Each node $v$ has an associated parameter $\lambda_v \in [0,1]$ which intuitively represents the willingness of an agent to change his or her habits. In this model, the obesity level of an agent evolves over time in response to the obesity levels of its neighbors to model the social nature of the adoption of habits that influence obesity. For a given agent $v$, we have \[x_v(t+1) = (1-\lambda_v)x_v(t) + \lambda_v\sum_{u \in \delta^{in}(v)} w_{(u,v)}(t+1)x_u(t)\] If we let $\overrightarrow{X(t)} \in \mathbb{R}^{|V|}$ be a vector denoting the obesity level of each node $v$, $\Lambda \in \mathbb{R}^{|V| x |V|}$ be the diagonal matrix formed by the parameters $\lambda_v$ such that $\Lambda_{ii} = \lambda_i$, $I_{|V|} \in \mathbb{R}^{|V| x |V|}$ be the identity matrix and $W_{t+1} \in \mathbb{R}^{|V| x |V|}$ be the transposed weighted adjacency matrix at time $t+1$ (in other words, $W_{t+1} = \{w_{(v,u)}(t+1)\}_{(u,v)}$), we can express the obesity level update more concisely as: \[\overrightarrow{X}(t+1) = (I_{|V|}-\Lambda) \overrightarrow{X}(t) + \Lambda W_{t+1} \overrightarrow{X}(t)\] \[\Rightarrow \overrightarrow{X}(t+1) = (I_{|V|}-\Lambda+\Lambda W_{t+1}) \overrightarrow{X}(t)\]

\subsection{Objective Function}

We will consider two different objective functions. Firstly, we will consider a terminal objective function given by \[f^{terminal}(\{A_t\}_{t=1}^T,\{R_t\}_{t=1}^T) = \sum_{v \in S}x_v(T) - \alpha\sum_{t=1}^T \beta^t |A_t|\] The first sum in this objective represents the final obesity level of individuals in the target group. The second term in the objective is a penalty for introducing additional edges into the graph. This models the fact that the public health provider will incur increasing costs (financial, operational or other) for introducing an increasing number of edges into the graph. $\beta$ is a parameter of the model that is greater than $1$. $\alpha$ is a model parameter that is between $0$ and $1$. Thus, the penalty for introducing a fixed number of edges into the graph at time $t$ increases as $t$ increases. Secondly, we will consider a cumulative objective function given by \[f^{cumulative}(\{A_t\}_{t=1}^T,\{R_t\}_{t=1}^T) = \sum_{v \in S} \sum_{t=1}^T x_v(t) - \alpha\sum_{t=1}^T \beta^t |A_t|\] Note that the second term in this objective function is the same penalty term that is included in the terminal objective function. The first term in this objective function represents the cumulative obesity level of individuals in the target group over the entire horizon. Thus in this case, an effective intervention should rapidly improve the health levels of target individuals (whereas in the terminal case, we only care about the final health level and not how quickly we arrive at that level). Thus, letting $S_0$ represent the initial state of the system (that is the initial graph, initial obesity levels and initial edge weights), we hope to solve the problem \[ \max_{\{A_t\}_{t=1}^T} \quad \mathbb{E}_{\{R_t\}_{t=1}^T}[f^*(\{A_t\}_{t=1}^T,\{R_t\}_{t=1}^T)|S_0] \] where $f^*$ is either the terminal objective or the cumulative objective.

\subsection{Policy}

A policy function $X^{\pi}$ uniquely specifies what action to take from any given state. In other words, $X^{\pi}$ is a function that takes a state $S_t$ as input and outputs an action $X_t$. The majority of the remainder of this work will explore various policies to tackle this problem.

\section{Previous Work}

The use of networks to model various aspects of human interaction is a very active quantitative research area that draws its foundation in part from the social sciences. Moreover, optimization problems are often posed on top of an underlying network structure to achieve a humanitarian goal. Several such problems fall under the umbrella of influence maximization, which broadly refers to the problem of selecting a set of nodes in a graph such that when a given diffusion process (which can be thought of as information spreading in a network) is started from the nodes, the total number of nodes reached by that process is maximized \cite{influence_max}. For example, \cite{group_fairness} investigates influence maximization under group fairness constraints to ensure that the benefits of a public health HIV intervention is fairly distributed across marginalized groups in the network. Similarly, \cite{homeless} explores influence maximization under unknown network dynamics to help non-profits to better spread HIV prevention information among homeless youth. Even outside of the broad category of influence maximization, most optimization problems over networks that have been studied extensively involve selecting some subset of nodes in the network. For instance, \cite{suicide} considers the problem of selecting a subset of nodes in a network to serve as peer monitors for suicide prevention where the performance of the chosen monitors is uncertain.

To our knowledge, \cite{Wilder} is the first work to explore an optimization problem over a graph that involves selecting edges rather than nodes and the accompanying algorithmic challenges. In the paper "Optimizing Network Structure for Preventative Health," Bryan Wilder and colleagues introduce a novel framework for modeling preventative health interventions as social network optimization problems which they coin the preventative health network intervention (PHNI) framework \cite{Wilder}. Although similar, the problem formulation presented in the previous section builds on the PHNI framework in three significant ways:

\begin{enumerate}
	\item In our model, each edge $(u,v)$ has a weight $w_{(u,v)}(t)$ associated with it for each time period $t$ denoting the extent to which agent $u$ influences agent $v$. The PHNI framework implicitly assumes that each neighbor of a given node influences that node equally and constantly over the course of the learning process.
	\item Our model includes a parameter $\lambda_v$ for each agent $v$ which models the fact that each agent will alter their behavior at varying rates. The PHNI framework includes a single parameter $\lambda$ used across all agents; agents are assumed to change their habits at the same rate.
	\item In our model, the social planner can choose to make changes to the network structure at the beginning of each time $t$. The PHNI formulation only allows the social planner to make changes to the network at $t=1$.
\end{enumerate}
\noindent
Introducing the edge weights $w_{(u,v)}(t)$ and the parameters $\lambda_v$ creates a learning model that more accurately reflects DeGroot Learning \cite{learning_framework} than the model employed in the PHNI framework by allowing each node in the network to update its state as a possibly different linear combination between its state at the previous time step and a weighted average of the states of its neighbors in the graph. The PHNI framework employs state updates that use a simple average of the states of the neighbors of a node in the graph. This latter approach ignores the nuance that individuals are not necessarily influenced to the same extent by different individuals in their social circle. Degroot Learning is a model for social learning that has been empirically validated to be a more accurate model for social learning than common alternatives such as bayesian models \cite{social_learning}. Lastly, allowing the social planner to make changes to the network at each time $t$ rather than only at time $t=1$ allows the health intervention to be dynamic and adaptive rather than static.

\section{Proposed Algorithms}

\subsection{Heuristic Myopic Policy}

We will first focus on the cumulative objective. Here, we will develop a Myopic policy to solve this problem. The structure of this problem (and most sequential decision problems) is such that a decision taken at time $t$ has a long-lasting impact on the system. A Myopic policy is a policy that ignores the long-term effect of making a decision at time $t$ and only considers that decision's immediate impact on the system \cite{RLSO}. Noticing that for fixed $t$, the immediate impact of decision $A_t$ is reflected in the state $S_{t+1}$, the exact Myopic policy is given by \[X^{M}(S_t) = \argmax_{A_t \in D_t} \quad \mathbb{E}_{R_t} \Bigg{[}\bigg{[}\sum_{v \in S} x_v(t+1) \bigg{]}- \alpha\beta^{t} |A_{t}|\Big{|}A_t \Bigg{]}\] We will often write expectations that have the above structure throughout this section. It is important to note that the second term in the expectation is a deterministic function of $A_t$, so the expectation is actually only over the first term. For a fixed node $v \in S$ and fixed $t$, $x_v(t+1)$ depends only on edges $(u, w) \in A_t$ that satisfy $w = v$. This is a direct consequence of the state update equation $x_v(t+1)$. Recall that \[x_v(t+1) = (1-\lambda_v)x_v(t) + \lambda_v\sum_{u \in \delta^{in}(v)} w_{(u,v)}(t+1)x_u(t)\] In other words, the only term in the objective affected by including an edge $(u,w)$ in the decision set $A_t$ is the term $x_w(t+1)$. This allows for a convenient decomposition of the Myopic policy. Letting $\overline{E_t}^v = \{(u, w): (u, w) \in \overline{E_t}, w=v\} \,\, \forall \, v \in S$, we have 
\begin{align*}
\max_{A_t \in D_t} \quad \mathbb{E}_{R_t} \Bigg{[}\bigg{[}\sum_{v \in S} x_v(t+1) \bigg{]}- \alpha\beta^{t} |A_{t}|\Big{|}A_t \Bigg{]} = \mathlarger{\sum_{v \in S}} \quad \max_{\substack{A_t^v \subseteq \overline{E_t}^v \\ |A_t^v| \leq |\delta^{in}(v)|}} \mathbb{E}_{R_t} \bigg{[}x_v(t+1) - \alpha\beta^{t} |A_{t}^v| \,\Big{|}A_t^v \bigg{]}
\end{align*} Moreover, we have \[X^{M}(S_t) = \bigcup_{v \in S} \quad \argmax_{\substack{A_t^v \subseteq \overline{E_t}^v \\ |A_t^v| \leq |\delta^{in}(v)|}} \mathbb{E}_{R_t} \bigg{[}x_v(t+1) - \alpha\beta^{t} |A_{t}^v| \,\Big{|}A_t^v \bigg{]}\] Let $f^v(A_t^v) = \mathbb{E}_{R_t} \bigg{[}x_v(t+1) - \alpha\beta^{t} |A_{t}^v| \,\Big{|}A_t^v \bigg{]}$. Since the system dynamics are known to the agent, in particular the distribution over the edge removals $R_t$, given a feasible set $A_t^v$, $f^v(A_t^v)$ can be computed exactly. If we can efficiently maximize $f^v(\cdot)$ for fixed $v$, then we can efficiently compute the Myopic Policy. Unfortunately, maximizing $f^v(A_t^v)$ subject to $A_t^v \subseteq \overline{E_t}^v$ and $|A_t^v| \leq |\delta^{in}(v)|$ is a combinatorial optimization problem with a structure (in part due to the edge weight dynamics) that does not easily allow for efficient solving.

Algorithm 2 is a heuristic method to approximate the exact Myopic policy. For each node $v \in S$, Algorithm 2 computes a score for each edge that can be added to the network which is equal to the difference between the Myopic objective value if only the edge being considered is added and the Myopic objective value if no edges are added. If an edge's score is negative, that edge is not further considered. Let $k = |\delta^{in}(v)|$. After evaluating the scores of all candidate edges, the edges with the $k$ largest scores are added to the decision set. If fewer than k edges had a positive score, all such edges are added to the decision set. From the model dynamics, it follows that this expression can be computed exactly as 
\begin{align*}   
   f^v&(\{(u, v)\}) - f^v(\emptyset) = x_u(t) \cdot \lambda_v \sum_{r \in \delta^{in}(v)} \frac{\mathbb{P}\Big{(}(r, v) \in R_t \,\Big{|}A_t^v = \{(u, v)\}\Big{)}}{\Gamma_r^v(t)} + \lambda_v \cdot\\
   & \sum_{r \in \delta^{in}(v)}\frac{\mathbb{P}\Big{(}(r, v) \in R_t \,\Big{|}A_t^v = \{(u, v)\}\Big{)}}{\Gamma_r^v(t)} \Big{[}|\delta^{in}(v)|^{2} \cdot \big{(}\Delta^v(t)-w_{(r, v)}(t)x_r(t)\big{)} \Big{]} - \lambda_v\Psi^v(t) -\alpha\beta^t \numberthis \label{eq:added1}
\end{align*} where we have $\Psi^v(t) = \sum_{w \in \delta^{in}(v)}\Big{[} \Big{(} (1-\mu)w_{(w,v)}(t)+\frac{\mu}{|\delta^{in}(v)|} \Big{)} x_w(t) \Big{]}$ and $\\ \Delta^v(t) = \sum_{w \in \delta^{in}(v)} w_{(w,v)}(t)x_w(t)$. The detailed derivation for \ref{eq:added1} is omitted for the purposes of brevity. Computing $f^v(\{(u, v)\})-f^v(\emptyset)$ for an arbitrary $(u, v) \in \overline{E_t}^v$ takes time proportional to $|\delta^{in}(v)|$. However, after making an evaluation $f^v(\{(u, v)\}) - f^v(\emptyset)$, a subsequent evaluation $f^v(\{(r, v)\}) - f^v(\emptyset)$ can be computed in constant time. The operation of selecting the $k$ largest edges is performed using a max heap - the runtime of this operation is $O(m+k\log(m))$ where $m$ is the number of edges being considered. The runtime of Algorithm 2 is given by
\begin{align*}
    \text{Runtime (Algorithm 2)} &= \sum_{v \in S}\Big{[}|\delta^{in}(v)| + |\overline{E_t}^v| \Big{]}+\sum_{v \in S} \Big{[} |\overline{E_t}^v| + |\delta^{in}(v)| \cdot \log(|\overline{E_t}^v|) \Big{]} \\
    &= 2|\overline{E_t}| + \sum_{v \in S} \Big{[} |\delta^{in}(v)| \cdot (\log(|\overline{E_t}^v|)+1) \Big{]} \\
    &\leq 2|\overline{E_t}| + \sum_{v \in S} \Big{[} |V| \cdot (\log(|\overline{E_t}|)+1) \Big{]} \\
    &\leq 2|\overline{E_t}| + |V|^2 \cdot (\log(|\overline{E_t}|)+1) \numberthis \label{eq:eq7}
\end{align*} where we use the number of nodes in the graph ($|V|$) to upper bound both the in-degree of any node ($|\delta^{in}(v)|$) and the number of nodes in the target set ($|S|$). Thus, the runtime of the Heuristic Myopic Policy is $O\big{(}|\overline{E_t}|+|V|^2\log(|\overline{E_t}|)\big{)}$.
\begin{algorithm}
\SetAlgoLined
\KwResult{$A_t$}
 $A_t \gets \emptyset$ \\
 \For{$v \in S$}{
  scores $\gets$ empty symbol table \\
  \For{$(u,v) \in \overline{E_t}^v$}{
    \If{$f^v(\{(u, v)\}) - f^v(\emptyset) > 0$}{
     scores[(u,v)] = $f^v(\{(u, v)\}) - f^v(\emptyset)$
    }
  }
  // Select the edges that produce the greatest improvement without \\
  // violating the decision constraints. \\
  $A_t \gets A_t \, \cup \, \delta^{in}(v)$ edges with the greatest value in scores
}
 \caption{Heuristic Myopic Policy}
\end{algorithm} 

\subsection{Heuristic Myopic Policy Optimality In A Simplified Environment}

It is important to remember that Algorithm 2 is simply an approximation of the true Myopic Policy. This motivates a natural question: Is Algorithm 2 a good approximation to the true Myopic Policy? Unfortunately, this question cannot be readily answered in large part due to the complex environment dynamics, particularly due to the very nonlinear edge weight dynamics. Consider instead a simplified environment that is identical to the environment and dynamics we have been considering thus far except that $w_{(u,v)}(t) = \frac{1}{|\delta^{in}(v)|} \, \forall \, v \in V$ and $\forall \, t$. Expressed in words, this is equivalent to considering an environment in which every node in the network is equally influenced by its neighbours at all times. Although our earlier question cannot be answered for the true environment, it is possible to answer it for this simplified environment. In this simplified environment, Algorithm 2 exactly computes the true Myopic Policy. Let $X^{HM}$ denote Algorithm 2. Recall that $X^{M}$ denotes the Myopic Policy. 

\begin{theorem}
If $w_{(u,v)}(t) = \frac{1}{|\delta^{in}(v)|} \, \forall \, v \in V$ and $\forall \, t$, then for any state $S_t$ we have $X^{HM}(S_t) = X^{M}(S_t)$ if $X^{M}(S_t)$ is unique.
\end{theorem}

\noindent
\begin{proof}

Let $w_{(u,v)}(t) = \frac{1}{|\delta^{in}(v)|} \, \forall \, v \in V$ and $\forall \, t$. Fix an arbitrary time $t$ and state $S_t$. Assume $X^{M}(S_t)$ is unique. Let $X_v^{myopic}(S_t) = \{(u,w) \in X^{M}(S_t): w = v\}$ and $X_v^{HM}(S_t) = \{(u,w) \in X^{HM}(S_t): w = v\}$. We have,  $X^{M}(S_t) = \cup_{v \in S} X_v^{myopic}(S_t)$ and $X^{HM}(S_t) = \cup_{v \in S} X_v^{HM}(S_t)$. Recall that by the decomposition from the previous section, \[X_v^{myopic}(S_t) = \argmax_{\substack{A_t^v \subseteq \overline{E_t}^v \\ |A_t^v| \leq |\delta^{in}(v)|}} \mathbb{E}_{R_t} \bigg{[}x_v(t+1) - \alpha\beta^{t} |A_{t}^v| \,\Big{|}A_t^v \bigg{]}\] We will show that $X_v^{myopic}(S_t) = X_v^{HM}(S_t) \, \forall \, v \in S$. Fix an arbitrary node $v \in S$. We will show that $X_v^{myopic}(S_t) \subseteq X_v^{HM}(S_t)$ and $X_v^{HM}(S_t) \subseteq X_v^{myopic}(S_t)$.

Let us first show that $X_v^{myopic}(S_t) \subseteq X_v^{HM}(S_t)$. For an arbitrary edge $(u,v) \in \overline{E_t}$, let $G_{(u,v)} = \{(w,v): (w,v) \in \overline{E_t}^v, f^v(\{(w,v)\}) > f^v(\{(u,v)\})\}$. By construction of Algorithm 2, we have $f^v(\{(u,v)\}) - f^v(\emptyset) > 0$ and $|G_{(u,v)}| \leq |\delta^{in}(v)| - 1$ if and only if we have $(u,v) \in X_v^{HM}(S_t)$. Thus, to show that $X_v^{myopic}(S_t) \subseteq X_v^{HM}(S_t)$, it suffices to show:

\begin{enumerate}
    \item $(u,v) \in X_v^{myopic}(S_t) \implies f^v(\{(u,v)\}) - f^v(\emptyset) > 0$ and
    \item $(u,v) \in X_v^{myopic}(S_t) \implies |G_{(u,v)}| \leq |\delta^{in}(v)| - 1$.
\end{enumerate} To show the first condition, note that for any set $A_t^v$ we have
\begin{align*}
    f^v(A_t^v) &= \mathbb{E}_{R_t} \bigg{[}x_v(t+1) - \alpha\beta^{t} |A_t^v| \,\Big{|}A_t^v\bigg{]} =(1-\lambda_v)x_v(t)+\frac{\lambda_v}{|\delta^{in}(v)|}\mathbb{E}_{R_t} \Bigg{[}\sum_{w \in \delta^{in}(v)} x_w(t) \,\Big{|}A_t^v \Bigg{]} -\alpha\beta^t|A_t^v| \\
    &= (1-\lambda_v)x_v(t)+\frac{\lambda_v}{|\delta^{in}(v)|} \Bigg{[}\sum_{u \in \delta^{in}(v)}x_u(t)+\sum_{w:(w, v) \in A_t^v} x_w(t) -\frac{|A_t^v|}{|\delta^{in}(v)|}\sum_{u \in \delta^{in}(v)}x_u(t) \Bigg{]}-\alpha\beta^t|A_t^v| \numberthis \label{eq:eq8}
\end{align*} Noting that $f^v(\emptyset) = (1-\lambda_v)x_v(t)+\frac{\lambda_v}{|\delta^{in}(v)|}\sum_{u \in \delta^{in}(v)} x_u(t)$, we have
\begin{align*}
    f^v(A_t^v) = f^v(\emptyset) + \frac{\lambda_v}{|\delta^{in}(v)|}\Bigg{[}\sum_{w:(w, v) \in A_t^v} x_w(t) - \frac{|A_t^v|}{|\delta^{in}(v)|}\sum_{u \in \delta^{in}(v)}x_u(t) \Bigg{]}-\alpha\beta^t|A_t^v| \numberthis \label{eq:eq9}
\end{align*} Consider an edge $(u,v) \in X_v^{myopic}(S_t)$. Evaluating equation \ref{eq:eq9} with $A_t^v = X_v^{myopic}(S_t)$ and $A_t^v = (X_v^{myopic}(S_t) \setminus \{(u,v)\})$ gives
\begin{align*}
    f^v(X_v^{myopic}(S_t))&-f^v(X_v^{myopic}(S_t) \setminus \{(u,v)\}) = \\
    & \frac{\lambda_v}{|\delta^{in}(v)|} \Bigg{[} \sum_{w:(w, v) \in X_v^{myopic}(S_t)} x_w(t) - \frac{|X_v^{myopic}(S_t)|}{|\delta^{in}(v)|}\sum_{w \in \delta^{in}(v)}x_w(t) \\
    & - \sum_{w:(w, v) \in (X_v^{myopic}(S_t) \setminus \{(u,v)\})} x_w(t) + \frac{|(X_v^{myopic}(S_t) \setminus \{(u,v)\})|}{|\delta^{in}(v)|}\sum_{w \in \delta^{in}(v)}x_w(t)\Bigg{]} \\
    & -\alpha\beta^t(|X_v^{myopic}(S_t)|-|(X_v^{myopic}(S_t) \setminus \{(u,v)\})|) \\
    =& \frac{\lambda_v}{|\delta^{in}(v)|}\Bigg{[}x_u(t)-\frac{1}{|\delta^{in}(v)|}\sum_{w \in \delta^{in}(v)}x_w(t)\Bigg{]} -\alpha\beta^t = f^v(\{(u,v)\})-f^v(\emptyset) \numberthis \label{eq:eq10}
\end{align*} By the uniqueness of $X_v^{myopic}(S_t)$, we must have $f^v(X_v^{myopic}(S_t))-f^v(X_v^{myopic}(S_t) \setminus \{(u,v)\}) > 0$. This combined with equation \ref{eq:eq10} implies $f^v(\{(u,v)\})-f^v(\emptyset) > 0$.

To show the second condition, consider again an edge $(u,v) \in X_v^{myopic}(S_t)$. Suppose $|G_{(u,v)}| > |\delta^{in}(v)| - 1$. This implies $|G_{(u,v)}| \geq |\delta^{in}(v)|$. Since $X_v^{myopic}(s_t)$ cannot violate the decision constraints, we must have $|X_v^{myopic}(S_t)| \leq |\delta^{in}(v)|$. This implies $|(X_v^{myopic} \setminus \{(u,v)\})| \leq |\delta^{in}(v)|-1$, which in turn implies $\exists (r,v) \in G_{(u,v)} \big{|} (r,v) \notin X_v^{myopic}(S_t)$. Evaluating equation \ref{eq:eq9} with $ \\ A_t^v = (X_v^{myopic}(S_t) \setminus \{(u,v)\})  \cup \{(r,v)\}$ and $A_t^v = X_v^{myopic}(S_t)$ gives
\begin{align*}
    f^v((&X_v^{myopic}(S_t) \setminus \{(u,v)\}) \cup \{(r,v)\})-f^v(X_v^{myopic}(S_t)) = \\
    & \frac{\lambda_v}{|\delta^{in}(v)|} \Bigg{[} \sum_{w:(w, v) \in (X_v^{myopic}(S_t) \setminus \{(u,v)\}) \cup \{(r,v)\}} x_w(t) - \frac{|(X_v^{myopic}(S_t) \setminus \{(u,v)\}) \cup \{(r,v)\}|}{|\delta^{in}(v)|}\sum_{w \in \delta^{in}(v)}x_w(t) \\
    & - \sum_{w:(w, v) \in X_v^{myopic}(S_t)} x_w(t) + \frac{|X_v^{myopic}(S_t)|}{|\delta^{in}(v)|}\sum_{w \in \delta^{in}(v)}x_w(t)\Bigg{]} \\
    & -\alpha\beta^t(|(X_v^{myopic}(S_t) \setminus \{(u,v)\}) \cup \{(r,v)\}|-|X_v^{myopic}(S_t)|) \\
    =& \frac{\lambda_v}{|\delta^{in}(v)|}\big{[}x_r(t)-x_u(t)\big{]} = f^v(\{(r,v)\})-f^v(\{(u,v)\}) \numberthis \label{eq:eq11}
\end{align*} 
\noindent
We know that $(r,v) \in G_{(u,v)}$. This implies $f^v(\{(r,v)\}) > f^v(\{(u,v)\})$, which in turn implies $f^v((X_v^{myopic}(S_t) \setminus \{(u,v)\}) \cup \{(r,v)\}) > f^v(X_v^{myopic}(S_t))$. But by the uniqueness of $X_v^{myopic}(S_t)$, we must have $f^v(X_v^{myopic}(S_t)) > f^v((X_v^{myopic}(S_t) \setminus \{(u,v)\}) \cup \{(r,v)\})$. This is a contradiction. Thus, we must have $|G_{(u,v)}| \leq |\delta^{in}(v)| - 1$. We have shown that $X_v^{myopic}(S_t) \subseteq X_v^{HM}(S_t)$.

It remains to show that $X_v^{HM}(S_t) \subseteq X_v^{myopic}(S_t)$. Consider an edge $(u,v) \in X_v^{HM}(S_t)$. We have $f^v(\{(u,v)\}) - f^v(\emptyset) > 0$ and $|G_{(u,v)}| \leq |\delta^{in}(v)| - 1$. Suppose $(u,v) \notin X_v^{myopic}(S_t)$. There are two possible cases: $|X_v^{myopic}(S_t)| < |\delta^{in}(v)|$ and $|X_v^{myopic}(S_t)| = |\delta^{in}(v)|$.

Case 1: If $|X_v^{myopic}(S_t)| < |\delta^{in}(v)|$, then the set $(X_v^{myopic}(S_t) \cup \{(u,v)\})$ is feasible. Evaluating equation \ref{eq:eq9} with $A_t^v = (X_v^{myopic}(S_t) \cup \{(u,v)\})$ and $A_t^v = X_v^{myopic}(S_t)$ gives
\begin{align*}
    f^v(&X_v^{myopic}(S_t) \cup \{(u,v)\})-f^v(X_v^{myopic}(S_t)) = \\
    & \frac{\lambda_v}{|\delta^{in}(v)|} \Bigg{[} \sum_{w:(w, v) \in X_v^{myopic}(S_t) \cup \{(u,v)\}} x_w(t)  - \frac{|X_v^{myopic}(S_t) \cup \{(u,v)\}|}{|\delta^{in}(v)|}\sum_{w \in \delta^{in}(v)}x_w(t) \\
    & - \sum_{w:(w, v) \in X_v^{myopic}(S_t)} x_w(t) + \frac{|X_v^{myopic}(S_t)|}{|\delta^{in}(v)|}\sum_{w \in \delta^{in}(v)}x_w(t)\Bigg{]}  -\alpha\beta^t(|X_v^{myopic}(S_t) \cup \{(u,v)\}|-|X_v^{myopic}(S_t)|) \\
    =& \frac{\lambda_v}{|\delta^{in}(v)|}\Bigg{[}x_u(t) - \frac{1}{|\delta^{in}(v)|}\sum_{w \in \delta^{in}(v)}x_w(t)\Bigg{]} - \alpha\beta^t = f^v(\{(u,v)\})-f^v(\emptyset) > 0 \numberthis \label{eq:eq12}
\end{align*} However, by definition of $X_v^{myopic}(S_t)$, we know $f^v(X_v^{myopic}(S_t) \cup \{(u,v)\}) - f^v(X_v^{myopic}(S_t)) < 0$. This is a contradiction.

Case 2: If $|X_v^{myopic}(S_t)| = |\delta^{in}(v)|$, then $|G_{(u,v)}| \leq |\delta^{in}(v)| - 1$ and $(u,v) \notin X_v^{myopic}(S_t)$ implies $\exists (r,v) \in X_v^{myopic}(S_t) \, \big{|} f^v(\{(u,v)\}) \geq f^v(\{(r,v)\})$. Evaluating equation \ref{eq:eq9} with $A_t^v = (X_v^{myopic}(S_t) \setminus \{(r,v)\}) \cup \{(u,v)\}$ and $A_t^v = X_v^{myopic}(S_t)$ gives
\begin{align*}
    f^v((X_v^{myopic}&(S_t) \setminus \{(r,v)\}) \cup \{(u,v)\})-f^v(X_v^{myopic}(S_t)) = \\
    & \frac{\lambda_v}{|\delta^{in}(v)|} \Bigg{[} \sum_{w:(w, v) \in (X_v^{myopic}(S_t) \setminus \{(r,v)\}) \cup \{(u,v)\}} x_w(t) - \frac{|(X_v^{myopic}(S_t) \setminus \{(r,v)\}) \cup \{(u,v)\}|}{|\delta^{in}(v)|}\sum_{w \in \delta^{in}(v)}x_w(t) \\
    & - \sum_{w:(w, v) \in X_v^{myopic}(S_t)} x_w(t) + \frac{|X_v^{myopic}(S_t)|}{|\delta^{in}(v)|}\sum_{w \in \delta^{in}(v)}x_w(t)\Bigg{]} \\
    & -\alpha\beta^t(|(X_v^{myopic}(S_t) \setminus \{(r,v)\}) \cup \{(u,v)\}|-|X_v^{myopic}(S_t)|) \\
    =& \frac{\lambda_v}{|\delta^{in}(v)|}\big{[}x_u(t)-x_r(t)\big{]}= f^v(\{(u,v)\})-f^v(\{(r,v)\}) \geq 0 \numberthis \label{eq:eq13}
\end{align*} By the uniqueness of $X_v^{myopic}(S_t)$, we must have $f^v((X_v^{myopic}(S_t) \setminus \{(r,v)\}) \cup \{(u,v)\})-f^v(X_v^{myopic}(S_t)) < 0$. This is again a contradiction. Since we have shown a contradiction in all possible cases under the assumption that $(u,v) \notin X_v^{myopic}(S_t)$, we must have $(u,v) \in X_v^{myopic}(S_t)$. This implies $X_v^{HM}(S_t) \subseteq X_v^{myopic}(S_t)$, which completes the proof.

\end{proof}

Note that although this theorem requires a unique solution $X^{M}(S_t)$, in the event that $X^{M}(S_t)$ is not unique we still have $f^v(X^{HM}(S_t)) = f^v(X^{M}(S_t))$. Although the proven equivalence does not directly translate into the true environment, it nevertheless suggests that Algorithm 2 should be a reasonable approach to the problem at hand. 

\subsection{Heuristic One Step Lookahead Policy}

The primary shortcoming of the Heuristic Myopic Policy is that it does not consider the significance of adding an edge to the graph beyond that edge's immediate impact on the objective function. This can potentially result in the Heuristic Myopic Policy making a decision that is immediately attractive but ultimately not as beneficial as other alternatives. A One Step Lookahead Policy is arguably the simplest policy that considers more than the immediate impact of making a decision. As its name suggests, a One Step Lookahead Policy takes the decision at time $t$ that maximizes the immediate value of the objective and the value of the objective at the following time step. Concretely, the exact One Step Lookahead Policy $X^{1LA}$ is given by 
\begin{align*}
    X^{1LA}(S_t) &= \argmax_{A_t \in D_t} \quad \mathbb{E}_{R_t, R_{t+1}} \Bigg{[} \sum_{t'=t}^{t+1}\Bigg{[}\bigg{[}\sum_{v \in S} x_v(t'+1) \bigg{]}- \alpha\beta^{t'} |A_{t'}|\Bigg{]}\Big{|}A_t, A_{t+1} = \emptyset \Bigg{]} \\
    &= \argmax_{A_t \in D_t} \quad \mathbb{E}_{R_t} \Bigg{[} \sum_{t'=t}^{t+1}\bigg{[}\sum_{v \in S} x_v(t'+1) \bigg{]}- \alpha\beta^{t} |A_{t}|\Big{|}A_t, A_{t+1} = \emptyset \Bigg{]}
\end{align*} Similar to the Myopic Policy, the system dynamics are such that the One Step Lookahead Policy can be decomposed as follows: \[X^{1LA}(S_t) = \bigcup_{v \in S} \quad \argmax_{\substack{A_t^v \subseteq \overline{E_t}^v \\ |A_t^v| \leq |\delta^{in}(v)|}} \mathbb{E}_{R_t} \bigg{[} \sum_{t'=t}^{t+1}x_v(t'+1) - \alpha\beta^{t} |A_{t}^v| \,\Big{|}A_t^v, A_{t+1}^v = \emptyset \bigg{]}\] Let $h^v(A_t^v) = \mathbb{E}_{R_t} \bigg{[} \sum_{t'=t}^{t+1}x_v(t'+1) - \alpha\beta^{t} |A_{t}^v| \,\Big{|}A_t^v, A_{t+1}^v = \emptyset \bigg{]}$. Here again, since the system dynamics are known to the agent, given a feasible set $A_t^v$, $h^v(A_t^v)$ can be computed exactly. Unfortunately, just as in the case of the Myopic Policy, optimizing $h^v(A_t^v)$ subject to $A_t^v \subseteq \overline{E_t}^v$ and $|A_t^v| \leq |\delta^{in}(v)|$ is a difficult combinatorial optimization problem that is not submodular. 
Algortihm 3 is a heuristic method to approximate $X^{1LA}$. Note its similarity to the Heuristic Myopic Policy. The two algorithms are identical except that the function evaluation $f^v(\{(u, v)\}) - f^v(\emptyset) > 0$ has been replaced by $h^v(\{(u, v)\}) - h^v(\emptyset) > 0$.

\begin{algorithm}
\SetAlgoLined
\KwResult{$A_t$}
 $A_t \gets \emptyset$ \\
 \For{$v \in S$}{
  scores $\gets$ empty symbol table \\
  \For{$(u,v) \in \overline{E_t}^v$}{
    \If{$h^v(\{(u, v)\}) - h^v(\emptyset) > 0$}{
     scores[(u,v)] = $h^v(\{(u, v)\}) - h^v(\emptyset)$
    }
  }
  // Select the edges that produce the greatest improvement without \\
  // violating the decision constraints. \\
  $A_t \gets A_t \, \cup \, \delta^{in}(v)$ edges with the greatest value in scores
}
 \caption{Heuristic One Step Lookahead Policy}
\end{algorithm} As was the case for the Heuristic Myopic Policy, to implement and determine the runtime of Algorithm 3 we must specify how to compute $h^v(\{(u, v)\}) - h^v(\emptyset)$. This expression can be computed as
\begin{align*}
    h^v(\{(u, v)\}) - &h^v(\emptyset) = (2-\lambda_v)\lambda_v \Bigg{[} x_u(t) \cdot \sum_{r \in \delta^{in}(v)} \frac{\mathbb{P}\Big{(}(r, v) \in R_t \,\Big{|}A_t^v = \{(u, v)\}\Big{)}}{\Gamma_r^v(t)} + \\
    & \sum_{r \in \delta^{in}(v)}\frac{\mathbb{P}\Big{(}(r, v) \in R_t \,\Big{|}A_t^v = \{(u, v)\}\Big{)}}{\Gamma_r^v(t)} \Big{[}|\delta^{in}(v)|^{2} \cdot \big{(}\Delta^v(t)-w_{(r, v)}(t)x_r(t)\big{)} \Big{]} \\
    & - \Psi^v(t) \Bigg{]} + \lambda_v \Bigg{[} x_u(t) \cdot \sum_{r \in \delta^{in}(v)} \frac{\mathbb{P}\Big{(}(r, v) \in R_t \,\Big{|}A_t^v = \{(u, v)\}\Big{)}}{\Gamma_r^v(t)} \cdot \\
    & \frac{(1-\mu)|\delta^{in}(v)|+\mu}{|\delta^{in}(v)|\Gamma_r^v(t)}+\sum_{r \in \delta^{in}(v)}\frac{\mathbb{P}\Big{(}(r, v) \in R_t \,\Big{|}A_t^v = \{(u, v)\}\Big{)}}{\Gamma_r^v(t)} \Bigg{[} \Theta^v(t)+\\
    & \Gamma_r^v(t)\Pi^v(t)-\bigg{[}(1-\mu)|\delta^{in}(v)|^2w_{(r,v)}(t)+\frac{\mu\Gamma_r^v(t)}{|\delta^{in}(v)|}\bigg{]}x_r(t+1) \Bigg{]} \Bigg{]}-\alpha\beta^t \numberthis \label{eq:eq19}
\end{align*}
where we have defined $\Theta^v(t) = (1-\mu)|\delta^{in}(v)|^2\sum_{w \in \delta^{in}(v)}w_{(w,v)}(t)x_w(t+1)$ and $\Pi_v(t) = \frac{\mu}{|\delta^{in}(v)|}\sum_{w \in \delta^{in}(v)}x_w(t+1)$. We omit the derivation of \ref{eq:eq19} for brevity.

Just as was the case for computing $f^v(\{(u, v)\})-f^v(\emptyset)$, computing $h^v(\{(u, v)\})-h^v(\emptyset)$ for an arbitrary $(u, v) \in \overline{E_t}^v$ takes time proportional to $|\delta^{in}(v)|$. However, after making an evaluation $h^v(\{(u, v)\}) - h^v(\emptyset)$, a subsequent evaluation $h^v(\{(r, v)\}) - h^v(\emptyset)$ can be computed in constant time. Since computing $h^v(\{(u, v)\})-h^v(\emptyset)$ and $f^v(\{(u, v)\})-f^v(\emptyset)$ are operations of the same complexity, the runtime of the Heuristic One Step Lookahead policy is $O\big{(}|\overline{E_t}|+|V|^2\log(|\overline{E_t}|)\big{)}$.

\subsection{Heuristic One Step Lookahead Policy Optimality In The Simplified Environment}

In section 4.2, it was shown that in a simplified environment that is identical to the true environment being considered except that $w_{(u,v)}(t) = \frac{1}{|\delta^{in}(v)|} \, \forall \, v \in V$ and $\forall \, t$, the Heuristic Myopic Policy exactly computes the true Myopic Policy. The same guarantee can be shown for the Heuristic One Step Lookahead Policy and the true One Step Lookahead Policy. Let $X^{H-1LA}$ denote the Heuristic One Step Lookahead Policy and recall that $X^{1LA}$ denotes the exact One Step Lookahead Policy.

\begin{theorem}
If $w_{(u,v)}(t) = \frac{1}{|\delta^{in}(v)|} \, \forall \, v \in V$ and $\forall \, t$, then for any state $S_t$ we have $X^{H-1LA}(S_t) = X^{1LA}(S_t)$ if $X^{1LA}(S_t)$ is unique.
\end{theorem}
\noindent
We omit the proof of Theorem 2 as it is identical in structure to the proof of Theorem 1. Similarly to Theorem 1, although Theorem 2 requires a unique solution $X^{1LA}(S_t)$, in the event that $X^{1LA}(S_t)$ is not unique we still have $h^v(X^{H-1LA}(S_t)) = h^v(X^{1LA}(S_t))$. This result suggests that Algorithm 3 is a reasonable method to approximate the exact One Step Lookahead Policy.

\subsection{Comparison Of The Heuristic Myopic Policy And The Heuristic One Step Lookahead Policy In The Simplified Environment}

Having constructed two different policies for the cumulative objective function, $X^{HM}$ and $X^{H-1LA}$, a natural question is raised: which of the two policies will perform better? We will explore this question analytically in this section and revisit it empirically in the results section. Intuitively, one might expect the Heuristic One Step Lookahead Policy to outperform the Heuristic Myopic Policy because the latter only considers the impact of making a decision at time $t$ while the former considers both a decision's immediate impact and part of its long term impact. Under certain conditions, this notion can be formalized concretely in the simplified model environment described previously. Let 
\begin{align*}
    C_t^v(A_t^v,\,A_{t+1}^v|S_t) &= \mathbb{E}_{R_t, R_{t+1}} \bigg{[} \sum_{t'=t}^{t+1}\Big{[}x_v(t'+1) - \alpha\beta^{t'} |A_{t'}^v|\Big{]} \,\Big{|}S_t, A_t^v, A_{t+1}^v \bigg{]} \numberthis \label{eq:eq26}
\end{align*} To allow for a fair comparison of the two policies, for fixed $S_t$ we compare the expected objective value after executing $X^{HM}$ at time $t$ and again at time $t+1$ to the expected objective value after executing $X^{H-1LA}$ at time $t$ and doing nothing at time $t+1$. Theorem 3 gives sufficient conditions in which the Heuristic One Step Lookahead policy outperforms the Heuristic Myopic policy. We stress that these conditions are not necessarily necessary conditions for the stated result.
\begin{theorem}
If $w_{(u,v)}(t) = \frac{1}{|\delta^{in}(v)|} \, \forall \, v \in V$ and $\forall \, t$, then for any state $S_t$, if either of the following conditions hold for a fixed $v \in V$, we have \[C_t^v(X^{H-1LA}(S_t),\, \emptyset | S_t) \geq C_t^v(X^{HM}(S_t),\, X^{HM}(S_{t+1})|S_t)\]
\begin{enumerate}
    \item $X^{HM}(S_{t+1}) = \emptyset$ 
    \item $X^{HM}(S_t) = \emptyset$ and $x_w(t) \geq \displaystyle\frac{1}{|\delta^{in}(v)|}\sum_{u \in \delta^{in}(v)}x_u(t) - \frac{\alpha\beta^t(\beta-1)|\delta^{in}(v)|}{(2-\lambda_v)\lambda_v} \,\, \forall \, w: (w,v) \in X^{HM}(S_{t+1})$
\end{enumerate}

\end{theorem}
\noindent
We have omitted the proof of Theorem 3 for brevity.

\subsection{Gradient Based Policy}

We will now focus on the terminal objective. In this section, we present a modified version of the RECONNECT algorithm developed in \cite{Wilder} that is applicable to our problem. We refer to this policy as the Gradient Based Policy. The Gradient Based Policy only makes a decision at time $t=0$. Concretely, the Gradient Based Policy, which we will denote by $X^{G}$, approximates the policy given by \[X^\pi(S_0) = \argmax_{A_0 \in D_0} \quad \mathbb{E}_{R_0} \Bigg{[}\sum_{v \in S} x_v(T) - \alpha |A_0|\Big{|}S_0, A_0, \{A_{t'}=\emptyset\}_{t'=1}^T \Bigg{]}\] $X^G$ assumes the simplified edge weight dynamics first introduced in section 4.2, specifically that $w_{(u,v)}(t) = \frac{1}{|\delta^{in}(v)|} \, \forall \, v \in V$ and $\forall \, t$. Let $W$ be the transposed weighted adjacency matrix. Suppose no action is taken at the initial time step. Then the final state vector is given by \[\overrightarrow{X}(T) = (\Lambda W+I-\Lambda)^T\overrightarrow{X}(0)\] where $\Lambda$ is a diagonal matrix with $\Lambda_{i,i}=\lambda_i$ and $I$ is the identity matrix. For fixed $A_0$ and $R_0$, let $Y(A_0,R_0)$ be the perturbation matrix with $Y_{v,u}=\frac{1}{|\delta^{in}(v)|} \,\,\forall (u,v) \in A_0$ and $Y_{v,u}=-\frac{1}{|\delta^{in}(v)|} \,\,\forall (u,v) \in R_0$. Note that with this notation, the updated weighted adjacency matrix after making decision $A_0$ and observing $R_0$ is given by $W+Y(A_0,R_0)$. Let $\Omega(Y(A_0,R_0)) = \Lambda \big{(}W+Y(A_0,R_0)\big{)}+I - \Lambda$ and let $\mathbf{1}_S$ be the column vector satisfying $\mathbf{1}_{S,u}=1$ if $u \in S$ and $0$ otherwise. Given this notation, the final objective value $f\big{(}Y(A_0,R_0)\big{)}$ after making decision $A_0$ and observing $R_0$ is given by $f\big{(}Y(A_0,R_0)\big{)}=\mathbf{1}_S^\top\Omega\big{(}Y(A_0,R_0)\big{)}^T\overrightarrow{X}(0)-\alpha|A_0|$. Thus, we seek a policy $X^G$ that approximates the policy given by \[X^\pi(S_0) = \argmax_{A_0 \in D_0} \quad \mathbb{E}_{R_0} \Bigg{[}\mathbf{1}_S^\top\Omega\big{(}Y(A_0,R_0)\big{)}^T\overrightarrow{X}(0)-\alpha|A_0| \Big{|}S_0\Bigg{]}\] 

\begin{algorithm}
\SetAlgoLined
\KwResult{$A_0$}
 $\nabla^0 \gets \nabla_Y\Big{[}\mathbf{1}_S^\top\Omega\big{(}Y(\emptyset,\emptyset)\big{)}^T\overrightarrow{X}(0)\Big{]} $ \\
 $A^0 \gets \argmax_{A_0 \in D_0} \mathbb{E}_{R_0}\Big{[}\langle Y(A_0,R_0), \nabla^0 \rangle\Big{]}$ \\
 \For{$l = \overline{1:L}$}{
  Sample $n$ sets of edge removals $\{r_i\}_{i=1}^n$ from $E_0(A^{l-1})$ \\
  $\Tilde{f}(A^{l-1}) \gets \frac{1}{n}\sum_{i=1}^n f\big{(}Y(A^{l-1},r_i)\big{)}$ \\
  $\Tilde{\nabla}^l \gets \frac{1}{n}\sum_{i=1}^n \nabla_Yf\big{(}Y(A^{l-1},r_i)\big{)}$ \\
  $A^l \gets \argmax_{A_0 \in D_0} \mathbb{E}_{R_0}\Big{[}\langle Y(A_0,R_0), \Tilde{\nabla}^l \rangle\Big{]}$ \\
} 
// Select the best sampled set of edge additions \\
$A_0 \gets \argmax_{l = \overline{0:L-1}}\Tilde{f}(A^l)$ \\
\caption{Gradient Based Policy}
\end{algorithm}

$X^G$ is given by Algorithm 4, an iterative gradient based optimization algorithm. In order to implement Algorithm 4, we must specify how to compute two quantities:
\begin{enumerate}
    \item $\nabla_Yf\big{(}Y(A,R)\big{)}$ for fixed $A$ and $R$
    \item $\argmax_{A_0 \in D_0} \mathbb{E}_{R_0}\Big{[}\langle Y(A_0,R_0), \Tilde{\nabla} \rangle\Big{]}$ for a given $\Tilde{\nabla}$
\end{enumerate}
\noindent
We start with the first quantity. For any $i \neq j$, we have $\Omega(Y)_{i,j} = \lambda_i(W_{i,j}+Y_{i,j})+1-\lambda_i$. We can ignore the case when $i = j$ because we cannot add self-loops to the network. Recalling the definition of $Y$, taking the derivative of $\Omega(Y)_{i,j}$ with respect to $Y_{i,j}$ gives $\diff{\Omega(Y)_{i,j}}{Y_{i,j}}=\frac{\lambda_i}{|\delta^{in}(i)|}$. Recall that $f\big{(}Y(A,R)\big{)} = \mathbf{1}_S^\top\Omega\big{(}Y(A_0,R_0)\big{)}^T\overrightarrow{X}(0)-\alpha|A_0|$. Using matrix calculus differentiation rules, we have \[\diff{f}{\Omega(Y)}=\sum_{r=0}^{T-1}\big{(}\Omega(Y)^r\big{)}^\top\mathbf{1}_S\overrightarrow{X}(0)^\top\big{(}\Omega(Y)^{T-1-r}\big{)}^\top\] Applying the chain rule, we have $\diff{f}{Y_{i,j}}=\frac{\lambda_i}{|\delta^{in}(i)|}\cdot\diff{f}{\Omega(Y)_{i,j}} - \mathbbm{1}_{\{Y_{i,j}=0\}}\alpha$. This entirely specifies how to compute $\nabla_Yf\big{(}Y(A,R)\big{)}$ for fixed $A$ and $R$. Computing $\diff{f}{\Omega(Y)}$ requires $T$ matrix multiplications if computed efficiently. Thus, computing a gradient evaluation $\nabla_Yf\big{(}Y(A,R)\big{)}$ has complexity $O(T|V|^w)$ where $w$ is the matrix multiplication constant which depends on the algorithm used for matrix multiplication. For all such known algorithms, $2.3 < w \leq 3$.

Moving on to the second quantity, note that we have
\begin{align*}
    \argmax_{A_0 \in D_0}\mathbb{E}_{R_0}\Big{[}\langle Y(A_0,R_0), \Tilde{\nabla} \rangle\Big{]} &=\argmax_{A_0 \in D_0}\sum_{(u,v)\in A_0}\Tilde{\nabla}_{v,u}-\mathbb{E}_{R_0}\Bigg{[}\sum_{(u,v)\in R_0}\Tilde{\nabla}_{v,u}\Bigg{]} \\
    &=\argmax_{A_0 \in D_0} \sum_{v \in S}\Bigg{[}\sum_{(u,v)\in A_0}\Tilde{\nabla}_{v,u}-\mathbb{E}_{R_0}\Bigg{[}\sum_{(u,v)\in R_0}\Tilde{\nabla}_{v,u}\Bigg{]}\Bigg{]} \numberthis \label{eq:eq36}
\end{align*} Equation \ref{eq:eq36} is very similar in structure to the optimization problem being solved by the Heuristic Myopic Policy in section 4.1. For similar reasons as those presented in sections 4.1 and 4.2, Algorithm 2 can be used to approximate the maximization problem given by \ref{eq:eq36} by letting \[f^v(\{(u,v)\})-f^v(\emptyset)=\Tilde{\nabla}_{v,u}-\sum_{r \in \delta^{in}(v)}\mathbb{P}\Big{(}(r, v) \in R_0 \,\Big{|}A_0^v = \{(u, v)\}\Big{)}\cdot\Tilde{\nabla}_{v,r}\] Executing Algorithm 2 with this modification does not change its complexity. Therefore, the complexity of computing an estimate for $\argmax_{A_0 \in D_0} \mathbb{E}_{R_0}\Big{[}\langle Y(A_0,R_0), \Tilde{\nabla} \rangle\Big{]}$ is $O\big{(}|\overline{E_0}|+|V|^2\log(|\overline{E_0}|)\big{)}$.

Computing the objective value $f\big{(}Y(A,R)\big{)}$ for fixed $A$ and $R$ requires raising the matrix $\Omega(Y)$ to the power $T$. If done efficiently, this operation has complexity $O(\log(T)|V|^w)$. Performing a runtime analysis of $X^G$ gives
\begin{align*}
    \text{Runtime (Algorithm 4)} &\leq C (L+1) n \Big{[}T|V|^w+|\overline{E_0}|+|V|^2\log(|\overline{E_0}|)+\log(T)|V|^w\Big{]} \\
    &=C (L+1) n \Big{[}\big{(}T+\log(T)\big{)}|V|^w+|\overline{E_0}|+|V|^2\log(|\overline{E_0}|)\Big{]} \numberthis \label{eq:eq37}
\end{align*} Where $C$ is some (possibly large) constant. Thus, the runtime of Algorithm 4 is $\\O\Big{(}nL\Big{[}\big{(}T+\log(T)\big{)}|V|^w+|\overline{E_0}|+|V|^2\log(|\overline{E_0}|)\Big{]}\Big{)}$. Recall that $2.3 < w \leq 3$.

\section{Experimental Setup}

The performance and runtime of the policies presented in this work are evaluated empirically through simulations. To conduct the desired simulations in the most realistic manner possible, we require access to data that will allow the construction of initial networks that reflect true social networks in Canada as best as possible. The 2015/2016 Canadian Community Health Survey Public Use Microdata File (PUMF) was used as the data source for network construction \cite{stats_can_1}. The 2015/2016 Canadian Community Health Survey PUMF contains data from health regions across Canada gathered by Statistics Canada from interviews with roughly 110 000 respondents aged 12 or older over a two-year period. The dataset contains detailed information about each respondent's geographic, demographic and health characteristics. Of the roughly 1300 features included in the dataset, we focus on the following 8 features: province, health region, gender, age, employment status, student status, adjusted BMI, and cultural/racial background.

\subsection{Network Construction}

Each agent in the network has a gender from the set \{Male, Female\}, an age from the set \{12-14, 15-17, 18-19, 20-24, 25-29, 30-34, 35-39, 40-44, 45-49, 50-54, 55-59, 60-64, 65-69, 70-74, 75-79, 80+\}, a race from the set \{white, non-white\}, a student status from the set \{yes, no\}, an employment status from the set \{yes, no, N/A\} and a health state from the set \{Non-overweight, Overweight, Obese\}. An employment status of N/A means that an individual is not employment eligible (for instance, the individual may be retired). For a fixed region (which could be the entire country, an individual province/territory or a single health region), $N$ agents are generated by sampling the aforementioned features according to empirically derived region specific distributions specified by the data. Let $A$ denote the set of agents (we have $|A| = N$). For a given set of agents $A$, let $A^O \subseteq A$ denote the set of agents who are obese. A set $S$ with $|S| = a \cdot |A^O|$ will be randomly selected from the set $A^O$ and a set $H$ with $|H| = b \cdot |A\setminus A^O|$ will be randomly selected from the set $A\setminus A^O$, the set of all non obese individuals. We fix $a = b = \frac{1}{2}$ in all the experiments we conduct. The value of the parameter $N$ varies across trials.

The Barabási-Albert graph generation model with spatial preferential attachment is used to generate the network after the agents have been sampled \cite{Wilder} \cite{graph_construction}. In this procedure, the first $m_0$ agents are placed into the graph and edges are added to make the graph fully connected. Formally, if we let $B$ denote the set of the first $m_0$ agents, then there exists edges $(u, v)$ and $(v, u) \; \forall \, u, v \in B$. The remaining agents arrive in the network one by one and form $m \leq m_0$ edges with existing nodes. The probability that an incoming node forms a link with a node already in the network is proportional to the degree of the existing node and the similarity between the demographic features of the existing node and the incoming node. Specifically, suppose agent $u$ arrives in the network while agent $v$ is already present. If $y_u$ is the feature vector for agent $u$, $y_v$ is the feature vector for agent $v$ and $d(v)$ is the in-degree of agent $v$, then agent $u$ forms a link with agent $v$ with probability proportional to $e^{-\rho||y_u-y_v||_2}d(i)$. Thus, arriving agents are more likely to form links with existing agents with high degrees and who have similar features to them. We fix $\rho = 0.1$ throughout our experiments. If a link is added between agents $u$ and $v$ in this process, we create a directed edge in both directions. Edge weights are initialized randomly (and independently) in the interval $[0,1]$ and normalized at the end of the network generating process such that we have $\sum_{u \in \delta^{in}(v)} w_{(u,v)} = 1$ $\forall \, v$.

\subsection{Policies Evaluated}

We empirically evaluate the performance of three baseline policies and the three policies introduced in section 4. The three baseline policies will be referred to as the Control Policy, the Initial Random Policy and the Perpetual Random Policy.
\begin{enumerate}
    \item \textbf{Control Policy:} At each time $t$, do nothing. This policy represents the absence of any intervention (no mentor-mentee pairs are ever created). This is useful to illustrate how the network evolves if no action is taken. The complexity of this policy is $O(T)$ because a constant operation is performed $T$ times.
    \item \textbf{Initial Random Policy:} At time $t=0$, for each healthy node $u \in H$ select one node randomly from the set $\{v \in S: (u,v) \notin E_t\}$, the set of target nodes that are not already influenced by $u$, and add an edge $(u,v)$ to $A_t$ provided that it does not violate the constraints on $A_t$. For times $t > 0$, do nothing. This policy represents an intervention that randomly assigns mentor-mentee pairs at the beginning of the time period and then never introduces additional pairings. The time complexity of this policy is $O(|V|)$.
    \item \textbf{Perpetual Random Policy:} At each time $t$, execute the Initial Random Policy as if $t=0$. This corresponds to an intervention that randomly assigns mentor-mentee pairs for every time period. This policy has complexity $O(T|V|)$.
    \item \textbf{Heuristic Myopic Policy:} At each time $t$, execute $X^{HM}$ (Algorithm 2). Note that $|\overline{E_0}| \geq |\overline{E_t}| \,\, \forall \, t$ because an equal number of edges are added and removed from the graph at each time $t$. Thus, the complexity of this policy is $\\O\big{(}T\big{[}|\overline{E_0}|+|V|^2\log(|\overline{E_0}|)\big{]}\big{)}$.
    \item \textbf{Heuristic One Step Lookahead Policy:} At each even time $t$, execute $X^{H-1LA}$ (Algorithm 3). At each odd time $t$, do nothing. The complexity of this policy is $O\big{(}T\big{[}|\overline{E_0}|+|V|^2\log(|\overline{E_0}|)\big{]}\big{)}$.
    \item \textbf{Gradient Based Policy:} At time $t=0$, execute $X^G$ (Algorithm 4) with $n=L=10$. For times $t > 0$, do nothing. This policy has complexity $\\O\Big{(}\big{(}T+\log(T)\big{)}|V|^w+|\overline{E_0}|+|V|^2\log(|\overline{E_0}|)\Big{)}$. Recall that $2.3 < w \leq 3$.
\end{enumerate}

\section{Experiment Results and Discussion}

\begin{figure*}[t]
  \includegraphics[width=\textwidth]{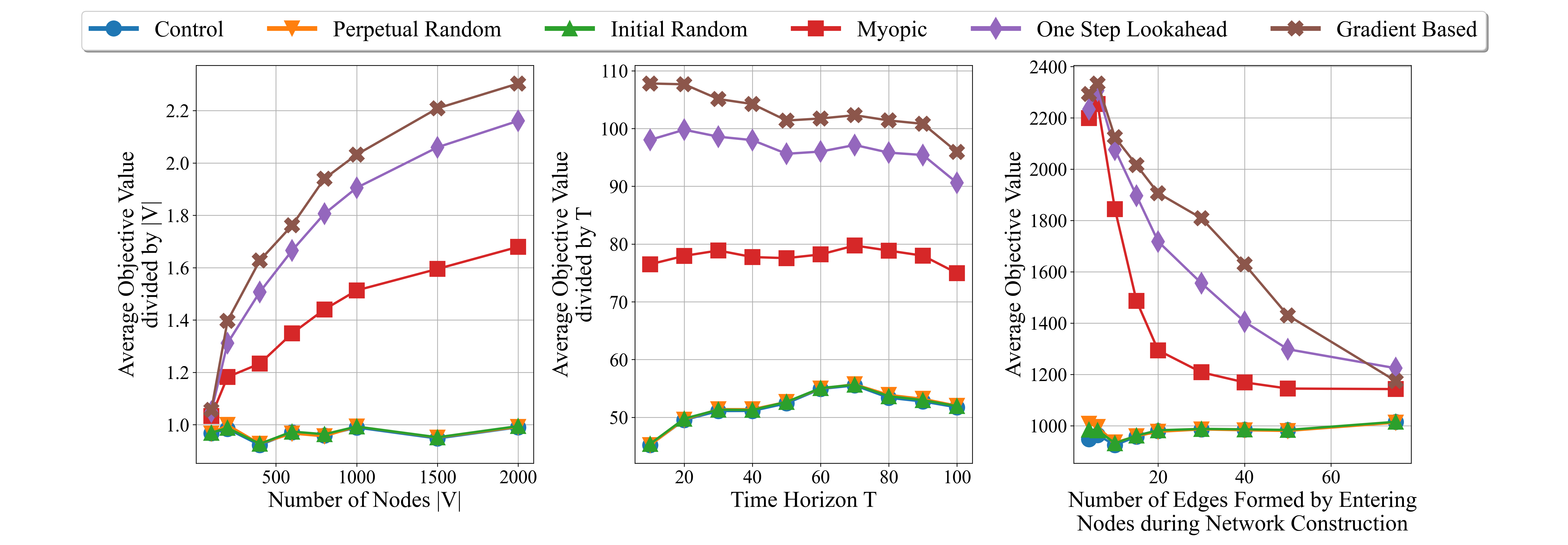}
  \caption{Normalized average objective value as a function of the number of nodes in the network (left), the time horizon of the simulation (center) and the number of edges formed by entering nodes during network construction (right).}
  \label{fig:obj_cumul_weight}
\end{figure*}

\begin{figure*}[t]
  \includegraphics[width=\textwidth]{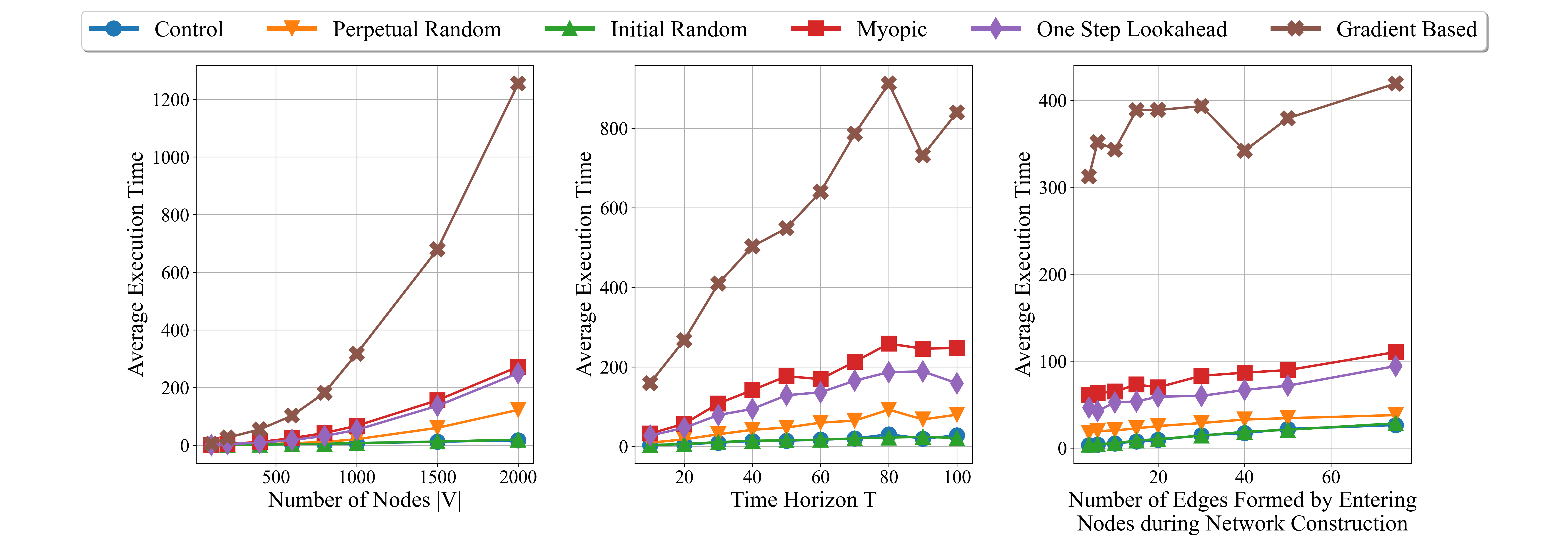}
  \caption{Average policy execution time as a function of the number of nodes in the network (left), the time horizon of the simulation (center) and the number of edges formed by entering nodes during network construction (right).}
  \label{fig:time_cumul_weight}
\end{figure*}

We focus on the experiment results from the cumulative objective trials to illustrate several key insights.  Figure \ref{fig:obj_cumul_weight} depicts the average objective value as a function of number of nodes, time horizon and the number of edges fromed by entering nodes during network construction. The first thing to note is that the Heuristic Myopic Policy, the Heuristic One Step Lookahead Policy and the Gradient Based Policy all significantly outperform the three baseline policies. Observe that the Gradient Based Policy outperformed all other policies across all experiment configurations with the exception of the configuration in which the number of edges formed by entering nodes during network construction was set to $75$. During this experiment configuration, the One Step Lookahead Policy achieved the greatest average objective value. It is important to note however that the difference in performance between the Gradient Based Policy and the One Step Lookahead Policy is often relatively small. The success of the Gradient Based Policy when used for the cumulative objective is particularly noteworthy because the Gradient Based Policy was designed specifically for the terminal objective and not for the cumulative objective. The One Step Lookahead Policy achieved a greater average objective value than the Myopic Policy across all trials. This is consistent with what we expected based on the intuition introduced in section 4.5. Moreover, this observation suggests that Theorem 3 can likely be generalized.

It is useful to investigate how each of the three experiment parameters we varied impacted the average objective value for the cumulative objective trials. Consider the plot that illustrates average objective value divided by $|V|$ as a function of $|V|$ (the number of nodes in the network). Notice that this plot is increasing for the three policies we developed. This means that these policies produce more favorable results on a per node basis as more nodes are introduced into the network. Focusing now on the plot of average objective value divided by $T$ versus $T$ in figure \ref{fig:obj_cumul_weight}, notice that this plot does not exhibit any consistent trend across all policies. The plot generally weakly increases for the baseline policies, is relatively constant for the Myopic Policy and is generally weakly decreasing for the One Step Lookahead and Gradient Based Policies. The fact that the plot is generally decreasing for the One Step Lookahead and Gradient Based Policies indicates that the marginal benefit of using either of these two policies decreases as the time horizon $T$ increases. Lastly, consider the plot of average objective value versus the number of edges formed by entering nodes during network construction ($m$). The parameter $m$ can be thought of as a measure of edge density in the network or equivalently as a measure of the average in-degree of a node. As $m$ increases, the performance of the Myopic, One Step Lookahead and Gradient Based Policies all decrease rapidly. This means that the public health intervention is becoming less effective as the density of the social network increases. This is not surprising. As an individual's social network increases in size, the influence that a single new addition can have on their behaviour correspondingly decreases. Thus, as the edge density in the network increases, one would expect a mentor-mentee based health intervention to have a lesser impact on the final health status of the network compared to the Control Policy.

Having investigated the empirical performance of the policies under the cumulative objective, we will now focus on the average execution time of the policies. Shifting focus to Figure \ref{fig:time_cumul_weight}, we see that the Gradient Based Policy took significantly longer to execute than all other policies across all trials. As expected based on the complexity of the policies, the three baseline policies had faster execution times than the policies we developed. Moreover, the Myopic and One Step Lookahead Policy empirically appear to have the same order of growth which is again consistent with the algorithmic complexity calculations we performed. The disparity in execution time between the Gradient Based Policy and the other policies is most pronounced in the plot of average execution time versus number of nodes $|V|$. Recall that the the Gradient Based Policy generally only marginally outperformed the One Step Lookahead Policy. Thus, although an intervention using the Gradient Based Policy would result in the greatest final health outcome of intervention participants among the policies tested, such an intervention does not scale to networks beyond a few thousand nodes. Conversely, an intervention using the One Step Lookahead Policy achieves a similar albeit slightly lower final health outcome but is able to scale to much larger networks. The plot of average execution time versus time horizon $T$ in Figure \ref{fig:time_cumul_weight} illustrates the rapid increase in runtime of the Gradient Based Policy as a function of $T$ compared to that of the One Step Lookahead Policy. This is consistent with the calculated complexities for each of these algorithms. The plot of average execution time versus the number of edges formed by an entering node ($m$) in Figure \ref{fig:time_cumul_weight} is gently increasing with $m$ for all policies. One might expect the opposite trend from looking at the calculated order of growth of each of the policies we developed. Recall that $|\overline{E_0}|$ denotes the initial number of edges that can potentially be formed between a node in the healthy set and a node in the target set that do not already exist in the graph. Thus, as the value of $m$ increases, the value of $|\overline{E_0}|$ decreases. Therefore, one might expect the runtime of the policies we developed to decrease as $m$ increases. However, it is important to recall that during the derivation of the runtime of the Myopic, One Step Lookahead and Gradient Based Policies, the number of nodes in the network $|V|$ was used to upperbound the in-degree $|\delta^{in}(v)|$ of any node $v$. Thus, the runtime of these policies could be rewritten more tightly as increasing functions of $|\delta^{in}(v)|$. The expected in-degree $|\delta^{in}(v)|$ of a node $v$ increases as $m$ increases and likely dominates the dependence of the runtime on $|\overline{E_0}|$, thereby resolving the apparent contradiction.

The above observations also held for the experiment results for the terminal objective trials. We note that the One Step Lookahead Policy and to a lesser extent the Myopic Policy still exhibited rather strong performance for the terminal objective despite having been developed specifically for the cumulative objective.

In summary, there are four major takeaways from the outcomes of the simulations:

\begin{enumerate}
    \item The Myopic Policy, the One Step Lookahead Policy and the Gradient Based Policy all significantly outperform the three baseline policies across all trials.
    \item The Gradient Based Policy outperforms all other policies in all trials with the exception of those for the cumulative objective where $m=75$. However, the Gradient Based Policy exhibits very poor scaling particularly in the number of nodes in the network.
    \item The One Step Lookahead Policy outperforms the Myopic Policy consistently across all trials.
    \item The One Step Lookahead Policy scales significantly better than the Gradient Based Policy while achieving comparable though slightly lower performance across all trials.
\end{enumerate}

\section{Conclusion}

In this work, we introduced a novel mathematical network model for preventative health interventions that allows for dynamic modification of the network and that more appropriately models how humans change their behavior in response to individuals they interact with when compared to existing alternatives. We developed the Heuristic Myopic Policy and the Heuristic One Step Lookahead Policy to approximately solve this novel formulation at large scale and we rigorously explored the theoretical properties of these algorithms. We proved analytically that under certain mild conditions, the One Step Lookahead Policy outperforms the Myopic Policy. To conclude our theoretical work, we developed a Gradient Based Policy inspired by the policy presented in \cite{Wilder}. Using demographic and health data obtained from the 2015/2016 Canadian Community Health Survey \cite{stats_can_1}, we created a realistic simulation environment for interventions occurring in the region of Montreal, Canada. We used this environment to empirically evaluate the performance of the Heuristic Myopic Policy, the Heuristic One Step Lookahead Policy and the Gradient Based Policy. We found that the Gradient Based Policy achieved the greatest objective function value across all trials. However, the Gradient Based Policy only marginally outperformed the Heuristic One Step Lookahead Policy despite having a significantly greater execution time. Consequently, for fixed computational resources, the Heuristic One Step Lookahead Policy can solve problems of greater magnitude than the Gradient Based Policy. In practice, assuming an intervention using the Heuristic One Step Lookahead Policy surpasses the accepted standard for quality of care, this means that with fixed resources the Heuristic One Step Lookahead Policy intervention can deliver good care to a far greater number of individuals diagnosed with obesity in Canada than an intervention based on the Gradient Based Policy.

Further work could build on the theoretical and empirical contributions presented in this work. Theorem 3 gives two sufficient conditions for which the Heuristic One Step Lookahead Policy outperforms the Heuristic Myopic Policy. Given that the Heuristic Lookahead Policy always outperformed the Heuristic Myopic Policy in our experiments, further theoretical work could attempt to analytically prove superiority of the Heuristic One Step Lookahead Policy in the general case. Future empirical work could concentrate on investigating the behaviour of the policies we developed for larger network sizes than those we experimented with. Evaluating the performance of the policies we developed using region specific data distributions different from those of the Montreal Region would also be useful.

\section*{Acknowledgments}
Special thanks to Professor Miklos Racz for his guidance and support while conducting this work.

\bibliographystyle{unsrtnat}
\bibliography{references}  

\begin{thebibliography}{13}
\providecommand{\natexlab}[1]{#1}
\providecommand{\url}[1]{\texttt{#1}}
\expandafter\ifx\csname urlstyle\endcsname\relax
  \providecommand{\doi}[1]{doi: #1}\else
  \providecommand{\doi}{doi: \begingroup \urlstyle{rm}\Url}\fi

\bibitem[Organization(2019)]{WHO}
World~Health Organization.
\newblock Health topics: Obesity, 2019.
\newblock URL \url{https://www.who.int/topics/obesity/en/}.

\bibitem[Boone et~al.(1977)Boone, Reilly, and Sashkin]{soc_learn}
Tim Boone, Anthony~J. Reilly, and Marshall Sashkin.
\newblock Social learning theory albert bandura englewood cliffs, n.j.:
  Prentice-hall, 1977. 247 pp., paperbound.
\newblock \emph{Group \& Organization Studies}, 2\penalty0 (3):\penalty0
  384--385, 1977.
\newblock \doi{10.1177/105960117700200317}.
\newblock URL \url{https://doi.org/10.1177/105960117700200317}.

\bibitem[Salvy(2016)]{change_behavior}
Sarah-Jeanne Salvy.
\newblock Home visitation programs: An untapped opportunity for the delivery of
  early childhood obesity prevention.
\newblock \emph{Obesity Reviews}, 18:\penalty0 149--163, 2016.

\bibitem[Canada(2020)]{stats_can_1}
Statistics Canada.
\newblock Canadian community health survey: Public use microdata file, 2020.
\newblock URL \url{https://www150.statcan.gc.ca/n1/en/catalogue/82M0013X}.

\bibitem[Wilder et~al.(2018)Wilder, Ou, de~la Haye, and Tambe]{Wilder}
Bryan Wilder, Han~Ching Ou, Kayla de~la Haye, and Milind Tambe.
\newblock Optimizing network structure for preventative health.
\newblock In \emph{Proceedings of the 17th International Conference on
  Autonomous Agents and MultiAgent Systems}, AAMAS '18, pages 841--849,
  Richland, SC, 2018. International Foundation for Autonomous Agents and
  Multiagent Systems.
\newblock URL \url{http://dl.acm.org/citation.cfm?id=3237383.3237507}.

\bibitem[Powell(2020)]{RLSO}
Warren~B. Powell.
\newblock \emph{Reinforcement Learning and Stochastic Optimization - A Unified
  Framework for Sequential Decisions}.
\newblock John Wiley and Sons, 2020.

\bibitem[Banerjee et~al.(2018)Banerjee, Jenamani, and Pratihar]{influence_max}
Suman Banerjee, Mamata Jenamani, and Dilip~Kumar Pratihar.
\newblock A survey on influence maximization in a social network.
\newblock \emph{CoRR}, abs/1808.05502, 2018.
\newblock URL \url{http://arxiv.org/abs/1808.05502}.

\bibitem[Tsang et~al.(2019)Tsang, Wilder, Rice, Tambe, and
  Zick]{group_fairness}
Alan Tsang, Bryan Wilder, Eric Rice, Milind Tambe, and Yair Zick.
\newblock Group-fairness in influence maximization.
\newblock \emph{CoRR}, abs/1903.00967, 2019.
\newblock URL \url{http://arxiv.org/abs/1903.00967}.

\bibitem[Wilder et~al.(2017)Wilder, Yadav, Immorlica, Rice, and
  Tambe]{homeless}
Bryan Wilder, Amulya Yadav, Nicole Immorlica, Eric Rice, and Milind Tambe.
\newblock Uncharted but not uninfluenced: Influence maximization with an
  uncertain network.
\newblock In Edmund Durfee, Michael Winikoff, Kate Larson, and Sanmay Das,
  editors, \emph{16th International Conference on Autonomous Agents and
  Multiagent Systems, AAMAS 2017}, Proceedings of the International Joint
  Conference on Autonomous Agents and Multiagent Systems, AAMAS, pages
  1305--1313. International Foundation for Autonomous Agents and Multiagent
  Systems (IFAAMAS), January 2017.
\newblock 16th International Conference on Autonomous Agents and Multiagent
  Systems, AAMAS 2017 ; Conference date: 08-05-2017 Through 12-05-2017.

\bibitem[Rahmattalabi et~al.(2019)Rahmattalabi, Vayanos, Fulginiti, and
  Tambe]{suicide}
Aida Rahmattalabi, Phebe Vayanos, Anthony Fulginiti, and Milind Tambe.
\newblock Robust peer-monitoring on graphs with an application to suicide
  prevention in social networks.
\newblock In \emph{Proceedings of the 18th International Conference on
  Autonomous Agents and MultiAgent Systems}, AAMAS ’19, page 2168–2170,
  Richland, SC, 2019. International Foundation for Autonomous Agents and
  Multiagent Systems.
\newblock ISBN 9781450363099.

\bibitem[DeGroot(1974)]{learning_framework}
Morris~H. DeGroot.
\newblock Reaching a consensus.
\newblock \emph{Journal of the American Statistical Association}, 69\penalty0
  (345):\penalty0 118--121, 1974.
\newblock ISSN 01621459.
\newblock URL \url{http://www.jstor.org/stable/2285509}.

\bibitem[Chandrasekhar et~al.(2015)Chandrasekhar, Larreguy, and
  Xandri]{social_learning}
Arun~G Chandrasekhar, Horacio Larreguy, and Juan~Pablo Xandri.
\newblock Testing models of social learning on networks: Evidence from a lab
  experiment in the field.
\newblock Working Paper 21468, National Bureau of Economic Research, August
  2015.
\newblock URL \url{http://www.nber.org/papers/w21468}.

\bibitem[Lu(2008)]{graph_construction}
Linyuan Lu.
\newblock Complex graphs and networks lecture 2: Generative models -
  preferential attachment schemes, 08 2008.
\newblock URL \url{http://people.math.sc.edu/lu/talks/lecture2.pdf}.

\end{thebibliography}






\end{document}